\documentclass[aip,graphicx]{revtex4-1}
\usepackage{graphicx}
\usepackage{dcolumn}
\usepackage{bm}

\draft 

\begin{document}

\title{Magnetic properties of epitaxial Fe$_3$O$_4$ films with various crystal orientations and TMR effect in room temperature} 

\author{Taro Nagahama}
\email[]{nagahama@eng.hokudai.ac.jp}
\author{Yuya Matsuda}
\author{Kazuya Tate}
\author{Shungo Hiratani}
\author{Yusuke Watanabe}
\author{Takashi Yanase}
\author{Toshihiro Shimada}

\affiliation{Graduate School of Engineering, Hokkaido University, Kita13 Nishi8, Kitak-ku, Sapporo, 060-8628, Japan}

\date{\today}

\begin{abstract}
Fe$_3$O$_4$ is a ferrimagnetic spinel ferrite that exhibits electric conductivity at room temperature (RT). Although the material has been predicted to be a half metal according to ab-initio calculations, magnetic tunnel junctions (MTJs) with Fe$_3$O$_4$ electrodes have demonstrated a small tunnel magnetoresistance effect. Not even the sign of the TMR ratio has been experimentally established. Here, we report on the magnetic properties of epitaxial Fe$_3$O$_4$ films with various crystal orientations. The films exhibited apparent crystal orientation dependence on hysteresis curves. In particular, Fe$_3$O$_4$(110) films exhibited in-plane uniaxial magnetic anisotropy. With respect to the squareness of hysteresis, Fe$_3$O$_4$ (111) demonstrated the largest squareness. Furthermore, we fabricated MTJs with Fe$_3$O$_4$(110) electrodes, and obtained an TMR effect of -12\% at RT. The negative TMR ratio corresponded to the negative spin polarization of Fe$_3$O$_4$ predicted from band calculations.
\end{abstract}


\maketitle 

Half metals that have 100\% spin polarization (P) at the Fermi level are key materials to fabricate spintronic devices because their high spin polarization enables very large magnetoresistance effects. The most impressive case is in magnetic tunnel junctions (MTJs) with epitaxial MgO tunnel barriers \cite{yuasa2004giant, parkin2004giant}. As transport in MgO-MTJs is dominated by coherent tunneling of $\Delta_1$ electrons with 100\% spin polarization, the TMR ratio has reached 600\% at RT \cite{ikeda2008tunnel}. Such a large TMR ratio has allowed us to fabricate highly functional spintronic devices like magnetoresistive random access memories (MRAMs). However, MTJs with MgO have stringent limitations where the crystal orientation should be bcc (001) due to band structure matching between MgO and the electrodes. Half metal is the solution to large TMR ratios without restricting the crystal structure or orientation. Thus far, many oxide materials have been proposed as candidates for half metals, e.g., CrO$_2$\cite{coey2002half}, La$_{0.7}$Sr$_{0.3}$MnO$_3$\cite{nadgorny2001origin}, and Fe$_3$O$_4$\cite{yanase1984band}. Of these materials, Fe$_3$O$_4$ has been considered to be the most promising as a half metal because of its high Curie temperature of 858 K, which is an advantage in applications to spintronic devices that require high Tc. The crystal structure is an inverse spinel with Fe$^{3+}$ cations occupying tetrahedral sites (A sites) and Fe$^{3+}$ and Fe$^{2+}$ cations occupying octahedral sites (B sites). The magnetic couplings between A and B sites are antiferromagnetic and the couplings at A-A or B-B are ferromagnetic; consequently, it is a ferrimagnetic material. As Fe$_3$O$_4$ exhibits good electric conductivity at RT due to the hopping of electrons between Fe$^{2+}$ and Fe$^{3+}$ on the B sites \cite{dionne2010magnetic}, the conduction electrons are 100\% spin polarized. As hopping is frozen on cooling, conductivity greatly decreases at low temperature, which is known as Verwey transition. The transition temperature, T$_v$, is ~120K \cite{verwey1939electronic}. The saturation magnetization of bulk Fe$_3$O$_4$ is ~510 emu/cc\cite{chikazumi2009physics}. According to Julliere's formula \cite{Julliere1975225}, MTJs with Fe$_3$O$_4$ electrodes are expected to exhibit very high TMR ratios due to large spin polarization. To date, researchers have fabricated MTJs with Fe$_3$O$_4$ and measured magnetoresistance; however, the TMR ratios have been small. Although the reason for this is not completely understood, such small TMR ratios can be attributed to imperfect antiparallel magnetic states in MTJs \cite{tiusan2001field}. The magnetization process of Fe$_3$O$_4$ films should be improved to achieve clear parallel and antiparallel magnetic configurations. We prepared epitaxial Fe$_3$O$_4$ films with various crystal orientations, and investigated their crystalline qualities and magnetic properties. We also fabricated MTJs with Fe$_3$O$_4$ electrodes and observed a negative TMR effect of -12\%.

The Fe$_3$O$_4$ thin films were prepared with three crystal orientations of (001), (110), and (111) by using a molecular beam epitaxy (MBE) system. The sample structures were: \\
1) an MgO(001) substrate/MgO (20 nm)/Fe$_3$O$_4$ (60 nm),\\
2) an MgO(110) substrate/MgO (20 nm)/Fe$_3$O$_4$ (60 nm), and\\
3) an Al$_2$O$_3$(0001) substrate/Pt (20 nm)/Fe$_3$O$_4$ (60 nm).\\
Following the deposition of MgO or Pt buffer layers, Fe$_3$O$_4$ thin film was formed by reactive deposition at a temperature (T$_{sub}$) of $300^\circ$C in an O$_2$ atmosphere of $4\times10^{-4}$ Pa. Then, the films were annealed at $600^\circ$C for 30 min in an O$_2$ atmosphere. The partial pressure of O$_2$ gas was $1\times10^{-4}$ Pa during annealing. All the samples were fabricated under the same growth conditions to enable the quality of Fe$_3$O$_4$ films to be compared. Epitaxial growth was observed with reflection high energy electron diffraction (RHEED) and the surface morphology was observed with atomic force microscopy (AFM). We also investigated the magnetization process at RT with a vibrating sample magnetometer (VSM).

Figs. 1 (a) and (b) show the RHEED patterns of Fe$_3$O$_4$(100) before and after O$_2$ annealing at $600^\circ$C for 30 min. The electron beam was incident along [100]. Fig. 1 (c) is an atomic force micrograph (AFM) of Fe$_3$O$_4$(100) after annealing. A streak RHEED pattern can be observed in Fig. 1 (a) meaning the Fe$_3$O$_4$ film grew epitaxially. In addition, p(1x1) surface reconstruction was observed \cite{chambers2000surface} \cite{tarrach1993atomic}. The streak pattern sharpened after annealing at $600^\circ$C in the O$_2$ atmosphere, as can be seen from Fig. 1 (b). A step-terrace structure can be confirmed from the AFM in Fig. 1 (c). The roughness average, R$_a$, was 0.12 nm, and the terrace width was ~200 nm.

Figs. 1 (d)-(f) show the RHEED patterns and AFMs of Fe$_3$O$_4$(110) grown on MgO(110). The incident electron beam direction was [-110]. A spotty pattern was obtained before annealing due to the island growth of Fe$_3$O$_4$ (110). However, the surface flatness was improved dramatically by O$_2$ annealing at $600^\circ$C, as can be seen from the RHEED pattern in Fig. 1 (e). The surface in the AFM of Fe$_3$O$_4$(110) after annealing in Fig. 1 (f) had anisotropic shapes along [100], which seemed to originate from the anisotropy of the MgO(110) substrate. R$_a$ was estimated to be 0.39 nm.

Finally, Figs. 1 (g)-(i) show RHEED patterns and AFMs of Fe$_3$O$_4$(111). The direction of the incident electron beam was [11-20]. Fig. 1(g) shows RHEED patterns of as-deposited Fe$_3$O$_4$(111). It shows streak patterns that indicate a flat surface and surface reconstruction. Terrace and step structures can be observed in the AFM of Fe$_3$O$_4$(111) after annealing in Fig. 1 (i); however, islands with a diameter of ~200 nm and height of several tens of nanometers were observed on the surface (not shown) in the AFM of a large area. The R$_a$ was estimated at 2.40 nm, which was one order of magnitude larger than the other crystal orientations. The large roughness could be attributed to the lattice mismatch between Fe$_3$O$_4$ and the Pt buffer layer\cite{Chambers2000105}. As the lattice constant of Fe$_3$O$_4$ was 0.8397 nm and that of MgO was 0.421 nm, the lattice mismatch was about 0.3\%. However, as the lattice constant of Pt was 0.392 nm, Fe$_3$O$_4$ lattice mismatch to the Pt buffer layer was 6.6\%. Such large lattice mismatch could give rise to a larger surface roughness for Fe$_3$O$_4$(111) than that for Fe$_3$O$_4$(100).

\begin{figure}
\includegraphics[width=8.5cm]{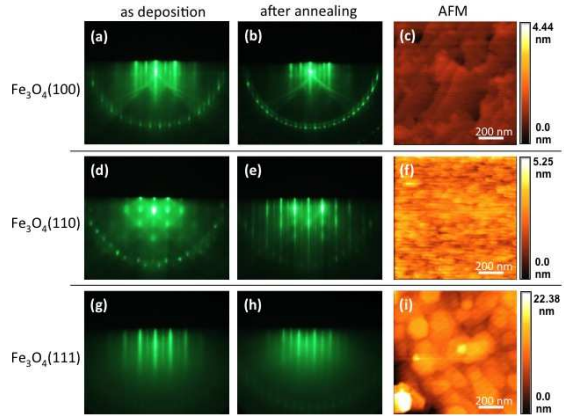}
\caption{\label{figure1} RHEED patterns and AFMs of epitaxial Fe$_3$O$_4$ (60 nm) films. RHEED patterns were taken after deposition at $300^\circ$C and annealing at $600^\circ$C. AFM observations were carried out after annealing. (a), (b), and (c) are for MgO(100)/Fe$_3$O$_4$(100) (60 nm). (d), (e), and (f) are for MgO(110)/Fe$_3$O$_4$(110) (60 nm). (g), (h), and (i) are for Al$_2$O$_3$(0001)/Pt(111) (20 nm)/Fe$_3$O$_4$(111) (60 nm).}
\end{figure}

The magnetization curves at RT for the Fe$_3$O$_4$ films are plotted in Fig. 2. The magnetic field was applied in plane. The diamagnetic components of the substrates were subtracted under the assumption that the magnetizations of the Fe$_3$O$_4$ were saturated at 5 kOe, which is the maximum field of VSM. The magnetization curve of Fe$_3$O$_4$ (100) is in Fig. 2 (a). The saturation magnetization (M$_s$) was 330 emu/cc, the remanent magnetization (M$_r$) was 100 emu/cc, and the coercive field (H$_c$) was 80 Oe. The remanent magnetization ratio (M$_r$/M$_s$) was 0.30. Fig. 2(b) plots the magnetization curves of Fe$_3$O$_4$ (110) where the directions of the magnetic field were [001] and [-110]. The saturation magnetization was 185 emu/cc for both magnetic field directions. M$_r$, H$_c$, and M$_r$/M$_s$ in the magnetic field along [001] were 30 emu/cc, 210 Oe, and 0.16, and those for [-110] were 100 emu/cc, 780 Oe, and 0.54. The magnetization process strongly depended on the directions of the magnetic field, viz., the squareness and M$_r$/M$_s$ were larger for the [-110] magnetic field than those for [100]. Nevertheless, the films had an anisotropic shape along the [100] direction, as shown in Fig. 1(f), and the films had a larger remanent ratio in the [-110] direction. Therefore, the anisotropy in Fig. 2 (b) was attributed to the magneto-crystalline anisotropy in Fe$_3$O$_4$. Saturation magnetization was 390 emu/cc, remanent magnetization was 290 emu/cc, and coercivity was 300 Oe in the magnetization curve of Fe$_3$O$_4$ (111). The remanent magnetization ratio was approximately 0.74, which was the largest value in the three crystal directions. The magnetic process was almost independent of the field directions. These values are summarized in Table 1. All the films exhibited smaller saturated magnetizations than the value for bulk Fe$_3$O$_4$ of 510 emu/cc. The reason for this is that the external field was not sufficient to saturate the magnetic moments in the Fe$_3$O$_4$ films. According to previous studies, Fe$_3$O$_4$ thin films contain considerable numbers of antiphase boundaries (APBs) \cite{margulies1997origin} that make the Fe$_3$O$_4$ hard to saturate magnetically due to antiferromagnetic coupling at the APBs.

\begin{figure}
\includegraphics[width=8.5cm]{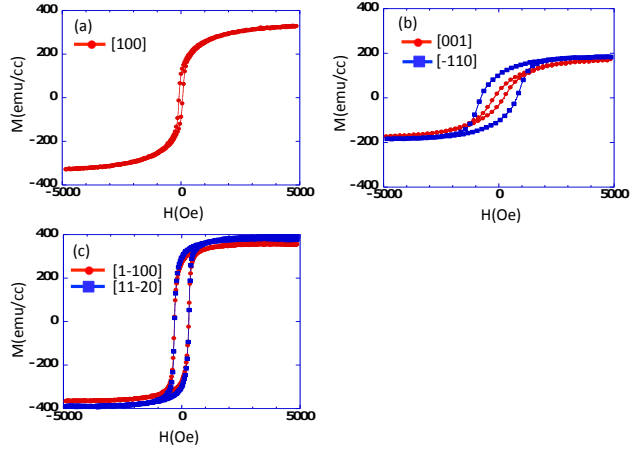}
\caption{\label{figure2} Hysteresis curves obtained from VSM measurements at RT for epitaxial Fe$_3$O$_4$ films with various crystal orientations. (a) is Fe$_3$O$_4$(100), (b) is Fe$_3$O$_4$(110), and (c) is Fe$_3$O$_4$(111). Directions of magnetic fields are given in plots.}
\end{figure}

\begin{table}
\caption{\label{tab:table4} Magnetic characteristics of Fe$_3$O$_4$ films with various crystal orientations. }
\begin{ruledtabular}
\begin{tabular}{lcccc}
 &Ms&Mr&Hc&Mr/Ms\\
 &(emu/cc)&(emu/cc)&(Oe)&\\
\hline
Fe$_3$O$_4$(100)&330&100&80&0.30\\
Fe$_3$O$_4$(110) H//[001]&185&30&210&0.16\\
Fe$_3$O$_4$(110) H//[-110]&185&100&780&0.54\\
Fe$_3$O$_4$(111)&390&290&300&0.74\\
\end{tabular}
\end{ruledtabular}
\end{table}

We fabricated the MTJs with Fe$_3$O$_4$(110) electrodes, and measured the tunnel magnetoresistance effect. The film structure was MgO(110)/NiO(110) (5 nm)/Fe$_3$O$_4$(110) (60 nm)/Al$_2$O$_3$ (2.4 nm)/Fe (5 nm)/Co (10 nm)/Au (30 nm). The NiO layer was inserted to suppress the diffusion of Mg from the substrates. Junctions of $10 \times 10 \mu m^2$ were fabricated by photolithography, Ar ion milling, and sputtering. The junctions demonstrated a clear TMR effect of -12\% at RT, as shown in Fig. 3. The negative MR agreed with the ab-initio calculations that predicted negative spin polarization in Fe$_3$O$_4$. To the best of our knowledge, these are the first experimental results of a negative MR ratio with an AlO barrier and Fe$_3$O$_4$ electrodes \cite{Aoshima:2003aa}\cite{Matsuda:2002aa}\cite{opel2011novel}\cite{seneor:1999aa}\cite{Park:2003aa}. The polarization of Fe$_3$O$_4$ deduced from the MR ratio based on Julliere's formula was -16\%, in which the polarization of Fe/Al$_2$O$_3$ was assumed to be 40\% \cite{meservey1994spin}. Although the polarization was much smaller than the predicted value, -16\% is of the same order as the reported values using various barrier materials \cite{hu2002negative}\cite{alldredge2006spin}. 

\begin{figure}
\includegraphics[width=8.5cm]{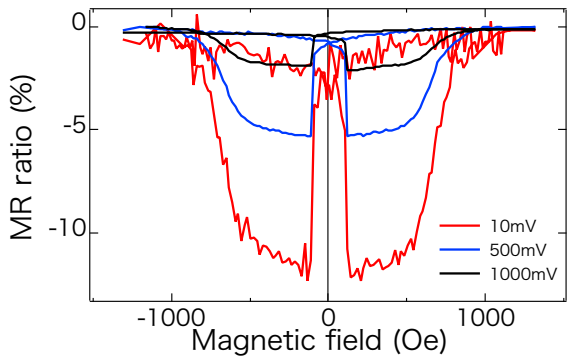}
\caption{\label{figure3} TMR curve for MTJ of MgO(110)/NiO(110) (5 nm)/Fe$_3$O$_4$(110) (60 nm)/Al$_2$O$_3$ (2.4 nm)/Fe (5 nm)/Co (10 nm)/Au (30 nm) at RT. Red, blue, and black lines are TMRs with bias voltages of 10, 500, and 1000 mV.}
\end{figure}

Fe$_3$O$_4$ epitaxial films with various crystal orientations were fabricated by reactive MBE and all the films grew epitaxially. The Fe$_3$O$_4$ (110) films exhibited clear uniaxial magnetic anisotropy that originated from crystal anisotropy. The squareness of the hysteresis curves strongly depended on the crystal orientation. A negative MR ratio of -12\% was observed in the MTJs with Fe$_3$O$_4$(110) electrodes. Although the absolute value was small, the negative MR agreed with the theoretical predictions. 

We would like to express our gratitude to Prof. Yamamoto's group for the cooperation in microfabrications. This work was supported by JSPS KAKENHI Grant-in-Aid for Young Scientists (A) Grant Number 23686006 and the Collaborative Research Program of Institute for Chemical Research, Kyoto University (grant 2014-75).

\bibliography{magnetite}

\end{document}